\documentclass[11pt,a4paper]{article}
\usepackage{amsfonts,amsmath,amssymb,amsthm}
\usepackage[utf8]{inputenc}
\usepackage[T1]{fontenc}
\usepackage[francais]{babel}
\usepackage{mathptmx}
\usepackage{graphicx}
\usepackage{subfig}
\usepackage{float}
\usepackage{subfloat}
\usepackage{ifthen}
\usepackage{amsmath,amssymb}
\usepackage{graphicx}
\usepackage{amssymb}
\usepackage{url}
\usepackage{amscd}
\usepackage{fancyhdr}
\usepackage{fancyheadings}
\newtheorem{prop}{Proposition}[section]

\newtheorem{theo}[prop]{Theorem}
\newtheorem{lem}[prop]{Lemma}
\newtheorem{defi}[prop]{Definition}
\newtheorem{rem}[prop]{Remark}

\title{Self-adjointness and spectrum of Stark operators on finite intervals}
\author{\textbf{H. NAJAR}$^a$\qquad  \textbf{ M. ZAHRI}$^b$\\ $~$\\
  \small D\'epartement de Math\'ematiques Facult\'e des Sciences de Monastir (Tunisia)\\ Laboratoire de
  \small recherche: Alg\`ebre G\'eom\'etrie et  Th\'eorie Spectrale, LR11ES53\\
  \small (Tunisia). Email: hatemnajar@yahoo.fr\\
  \small $^b$ Email:zahri@gmx.net}
\begin{document}
\maketitle

\abstract{In this paper, we study self-adjointness  and spectrum of operators of the form $$H=\displaystyle -\frac{d^2}{dx^2}+Fx, F>0 \quad\text{on} \quad \mathcal{H}=L^{2}(-L,L).$$ $H$ is called Stark operator and describes a quantum particle in a quantum asymmetric well. Most of known results on mathematical physics does not take in consideration the self-adjointness and the operating domains of such operators. We focus on this point and give the parametrization of all self-adjoint extensions. This relates on self-adjoint domains of singular symmetric differential operators. For some of these extensions, we numerically, give the spectral properties of $H$. One of these examples performs the interesting phenomenon of splitting of degenerate eigenvalues. This is done using the a combination of the Bisection and Newton methods with a numerical accuracy less than $10^{-8}$.}
\\ $~$
\large
\linespread{1.2}
\section{Introduction} The study of self-adjoint domains of symmetric differential operators on Hilbert spaces is a central problem in the theory of partial differential operators. It has a deep background in mathematical physics. As it is already mentioned in \cite{Rob}, a lot of works confuse between the symmetry (Hermiticity) and the self-adjointness, which leads to a non precise and even incomplete results. Frequently the basic distinction between unbounded and bounded operators is not considered, or often it is neglected. For getting a self-adjoint operator, the symmetric condition for unbounded operator is in general not sufficient. For differential operators on bounded domains the boundaries conditions at the end of the interval can change the situation and the question of self-adjointness it becomes more subtle and there are several scenarios, and the presence of boundaries changes significantly the picture. The operator can be self-adjoint, essentially self-adjoint, having many self-adjoint extensions or no self-adjoint extension at all or even non symmetric. So in this situation, we need more analysis to characterize a truly self-adjoint operator. In fact, if an observable of a quantum system is constructed starting from a symmetric operator will not be able to give the result performing measurement of such observable until we have made precise which self-adjoint extension of the system operator represents the observable. In particular, the role of the boundary conditions that lead to self-adjoint operators is missing in some of the available quantum textbooks. \par
One could ask, why the self-adjoint property is important? To convince the readers of the importance of this subject, we perform two fundamental reasons:
\begin{itemize}
  \item It is well-known
 that in mathematical physics problems we are interested in, are observable with real spectrum, which are guaranteed only for self adjoint operators. Thus it is very important to know if the domains under which theses operators are self-adjoint. Outside theses domains, eigenvalues are not only real-valued, and hence the operators cannot be considered as physical operators.
  \item Every self-adjoint operator is the generator of a unitary group.
Indeed when $H$ is a self-adjoint operator, the operators
$$U_t = \exp(itH),\quad t\in \mathbb{R};$$
are a (strongly continuous one-parameter) unitary group, and $H$ is its generator \cite{ReSi}.
More generally an operator generates a unitary group if and only if it is self-adjoint. This is could be related to the existence of dynamics in quantum mechanics.
Not only the dynamical evolution is affected by the determination of the boundary conditions, or the self-adjoint extensions of
families of symmetric operators, but also the results of the measures realized on the system and also the
measurable quantities of the system \cite{ibo}.
\end{itemize}
\subsection{Stark operators on finite intervals}
Experimentally, the atomic Stark effect means the shifts viewed in atomic emission
spectra after placing the particle in a constant electric field of strength $F$. J. Stark, in the non-relativistic
quantum theory, this Stark effect is usually modeled by an
Hamiltonian operator that (in appropriately scaled units and with the atomic units $2m =
h =q= 1$ to simplify the equation) has the form
\begin{equation}
H(F) = -\Delta + V(x)+Fx.\label{eq1}
\end{equation}  Hamiltonian that is parameterized by operators of the form (\ref{eq1})
has been intensively studied in the last five decinies \cite{Mil,olen} and references therein.
A quantum well is a particular kind of structure in which one layer is
surrounded by two barrier layers. Theses layers, in which particles are confined, could be so thin that we
cannot neglect the fact that particles are waves. In fact, the allowed states in
this structure correspond to standing waves in the direction perpendicular to the layers. Mathematically this corresponds to the study of (\ref{eq1}) on a $L^2(I)$, with $I$ is a finite interval seen as the support of $V$ (known as the quantum well).
Basic properties of a quantum well could be studied through the simple particle in a box
model. In the case of infinite quantum well it is expected that the energy levels are quadratically spaced, the energy level spacing becomes large for narrow wells.\\
 When electric field is applied to quantum wells, their optical absorption spectrum near to
the band-gap energy can be changed considerably \cite{Mil,olen}, an effect known as
electro-absorption.
 The correct
results in that case can also be obtained by explicit expansion of the exact eigenvalue
condition, but this requires knowing the properties of the boundary behavior and self-adjointness and the domain of the operator which is forgotten in many physical papers. This is the subject of the present paper. In the following section we summarize related results and give a brief survey of literature.
\subsection{Self-adjointness}
 The analysis of self-adjointness and the role of the boundary in quantum systems has became a recent focus of activity in different branches \cite{bal} and references therein. Everitt  in \cite{Eve} gave some results on self-adjoint domains under the assumption of the limit circle case and the limit point case. Using the Glazman-Krein-Naimark theory Everitt et al. in \cite{Eve2} showed that there exists a one-to-one
correspondence between the set of all the self-adjoint extensions of a minimal operator generated
by quasi-differential expressions and the set of all the complete Lagrangian subspaces of
a related boundary space. They used symplectic geometry. Sun et al \cite{Sun1,Sun2,Sun3} presented a complete and cordial characterization of all self-adjoint extensions
of symmetric differential operators. This is done by giving a new decomposition of the maximal operator domain. The later result is generalized by Evans and Ibrahim in \cite{Eva}.  Fu in \cite{Fu} gives the characterizations of self-adjoint domains for singular symmetric operators by describing boundary conditions of domain of conjugate differential operator, with singular points. Most of These operators are defined on a weighted Hilbert
function space. Self-adjointness of momentum operators in generalized coordinates is given in \cite{Dom} and of curl operator in \cite{Hip}.
\newline  In \cite{ver}, the authors consider a Schr\"odinger operator with a magnetic field and no electric field on  a domain in the Euclidean
 space with compact boundary. They give sufficient conditions on the behavior of the magnetic field near the boundary which guarantees essential self-adjointness of the operator. The problem concerning scalar potentials is first studied in  \cite{kal,sim}, under sum assumption on the behavior of $V$ when approaching the boundary. In \cite{grb}, a characterization was given for symmetric even order elliptic operators in bounded regular domains. Recently in \cite{fac}, the authors established a bijection between the self-adjoint of the Laplace operator on bounded regular domain and the unitary operator on the boundary. In \cite{Kat}, Katsnelson consider the formal prolate spheroid differential operator on a finite symmetric interval and describes the self-adjoint boundary conditions. He proves that among all self-adjoint extensions there is a unique realization which lead to an operator commuting with the Fourier operator trounced on the considered interval. In \cite{Bon}, a self-adjoint extensions is considered for analyzing momentum and Laplace operator.  \newline
The current result deals with self-adjointness of Stark operator on finite intervals. We give all parametrization giving self-adjointness. It should be stressed that the self-adjoint extensions of different domains are parameterized by  a  unitary group. The result is based on the von Neumann theory which provides necessary and sufficient conditions for the existence of self-adjoint
extensions of closed symmetric operators in Hilbert space \cite{von}. This theory is fully general and complectly solve the problem of self-adjoint extensions of every densely defined and closed symmetric operator in abstract Hilbert space using unitary operator between each deficiency \cite{oli}. However, for specific classes like stark operators, it would be suitable to have a more concrete characterization of self-adjoint extension.
  At section $4$ some particular extensions are considered and spectral properties are given. We focus that, changing the self-adjoint extensions leads to different spectral results. The phenomena of splitting of degenerates eigenvalues is observed for some particular self-adjoint boundary conditions. In Section \ref{split}, we give more details on the subject of splitting phenomena. The energy spectrum is calculated using the stable-state schrodinger characterized by the specific potential structure with the constant electric field. The problem is solved by representing the eigenfunction as a superposition of the airy functions despite the potential is simple (square-well), which lead to an analytical solution for the equation transforming on a very computationally complexity. The problem overcame in many papers using perturbation method, we refer to \cite{Rob} among others. This has the disadvantages (as any perturbation method) to be effective only for weak-enough electric field and produce invalid results for strong ones. For this, developing numerical methods to calculate the energy spectrum remains pertinent. It should be stressed that in one of our studied examples, our calculation confirm the experimental results (splitting phenomena). This method could lead to an exactly solvable approximate model.
\section{The model and the result}
We consider the Stark operator
\begin{equation}H=-\frac{d^2}{dx^2}+Fx, F>0.\label{eqz}
\end{equation}
The functional space is the  Hilbert space $\displaystyle L^{2}([-L,L])$ equipped with the scalar product
 $$\langle f,g\rangle=\int_{-L}^Lf(x)\overline{g(x)}dx,\forall f,g\in L^{2}([-L,L]).$$
 We notice that (\ref{eqz}) corresponds to the (\ref{eq1}), with $V(x)=0$ if $x\in [-L,L]$ and $V(x)=+\infty$  if not.\\
  This model corresponds to a particle in a box : $[-L, L]$ for $L>0$ in a presence of an electric fled of strength $F$. From a mathematical standpoint
the situation could be seen to be quite similar to the one with the free Laplacian, but up to our knowledge, it did not appear before in the literature. So we
give details below.\\ The maximal domain in which $H$ is well defined will denote by $\mathcal{D}_{max}$, i.e$$ \mathcal{D}_{max}=\{f\in L^{2}([-L,L]); Hf\in L^{2}([-L,L])\}.$$ Consider the domain
$$\mathcal{D}_0=\{\psi\in \mathcal{D}_{max}\ {\rm{and}}, \psi(-L)=\psi(L)=0=\psi'(-L)=\psi'(L)\}.$$
It is a closed and densely defined operator. The density of $\mathcal{D}_0$ follows from the fact that $C_0^\infty([-L,L])\subset\mathcal{D}_0$. The closeness of $H$ is due to the fact that the maximal domain is considered. Moreover, using the density of $H^2([-L,L])$ in $L^2$, we can get the closeness property. Using integration by part we get that $H$ is also a symmetric operator. The adjoint of $H$ is
$H^*=H$ and $$\mathcal{D}(H^*)=\{\psi\in \mathcal{D}_{max}\ {\rm{withount \ any \ other \ condition}}\}.$$
Hence, $H$ is not a self-adjoint operator on $\mathcal{D}_0$ and the considered  domain is too small to be associated to aselfadjoint operator. So, $H$ does not represent any physical observable and can not generate any physical dynamics. \newline
Let us notice that in the case of bounded operators as the domain of a densely defined bounded operator can always be extended to the entire vector space, therefore, a bounded Hermitian operator is also self-adjoint. However, in the unbounded case the situation is pathological and a little bit more subtle.\newline
Thus, we are interested clarifying conditions and domain under which symmetric, densely-defined $H$ can be self-adjoint and to know its self-adjoint extensions.
\begin{theo}\label{th1}
Let $H$ be the operator defined by (\ref{eqz}). Then $H$ has infinitely many self-adjoint extensions, these possible self-adjoint extensions of $H$ are parameterized by a unitary
matrix $U\in U(2)$. Let us denote them by $H_U = (H, \mathcal{D}(U))$, here $\mathcal{D}(U)$ is the space of functions
$\phi \in  \mathcal{D}_{max}$ satisfying the following boundary conditions
\begin{equation}\begin{pmatrix}
   L\phi'(-L)- i\phi(-L)\\
   L\phi'(L)+i\phi(L)
\end{pmatrix}=U \begin{pmatrix}
   L\phi'(-L)+i\phi(-L)\\
   L\phi'(L)-i\phi(L)
\end{pmatrix}.
\end{equation}
Each self-adjoint extension has purely discrete spectrum.
\end{theo}The result of the last theorem could be related to the von Neumann theorem \cite{von} which a powerful tool used in such situation. The proof of Theorem \ref{th1} is given in the next section.
\section{Deficiency indices, von Neumann's theorem and self-adjoitness}
 First we recall the definition and some properties of deficiency indices. For a Hilbert space $\mathcal{H}$, and  operator $(A,\mathcal{D}(A))$ defined on $\mathcal{H}$, with $\mathcal{D}$ a dense subspace of $\mathcal{D}$. The domain $\mathcal{D}(A^*)$, of the
 adjoint $A^*$, is the space of functions $\varphi$ such that the linear form
$$\psi\to \langle A\varphi, \psi \rangle, $$
is continuous for the norm of $\mathcal{H}$. So there exists a $\psi^*\in \mathcal{H}$ such that
$$\langle A\varphi,\psi\rangle=\langle \varphi,\psi^*\rangle.$$
We define the adjoint $A^*$ by $\displaystyle A^*\psi=\psi^*$ \cite{ReSi4}.
The space $\mathcal{E}=\mathcal{D}(A^*)/\mathcal{D}(A)$. Is called factor space.
\begin{defi}
For a densely defined, symmetric and closed operator $(A,\mathcal{D}(A))$, we define the deficiency subspaces $\mathcal{D}_{\pm}$ by
$$\mathcal{D}_+=\{ \varphi\in \mathcal{D}(A^*), A^*\varphi=z_+\varphi, Im z_+>0\},$$
$$
\mathcal{D}_-=\{ \varphi\in \mathcal{D}(A^*), A^*\varphi=z_-\varphi, Im z_-<0\},
$$
with respective dimensions $d_+,d_-$. These are called the deficiency indices of the operator $A$ and will be denoted by the ordered pair $(d_+,d_-)$.
 \end{defi}
 We note that $d_+$ and $d_-$ are independent of the points $z_+$ and $z_-$ respectively \cite{Ak,dev,von}, so for simplicity we take $z_+=i$ and $z_-=-i$. The theorem below known as von Neumann theorem relates the deficiency indices to the number of self-adjoint extension of an operator for the proof see \cite{Ak, dev,von}.
\begin{theo}\label{neu}\cite{von}
For a symmetric and closed operator $A$ with deficiency indices $(d_+,d_-)$ there are three possibilities:
\begin{enumerate}
\item  If $d_+=d_-=0$, then $A$ is selfadjoint.
\item  If $d_+=d_-=d\geq 1$, then $A$ has infinitely many self-adjoint extensions, parameterized by a unitary $d\times d$ matrix.
\item If $d_+\neq d_-$, then $A$ has no selfadjoint extension.
\item The dimension of the factor space is $d_-+d-+$.
\end{enumerate}
\end{theo}
\begin{rem}
\begin{enumerate}
\item The first point of the last theorem is a necessary and sufficient condition.
\item The second point says that the set of all selfadjoint extensions is parameterized by $d^2$ real parameter.
\item The von Neumann's argument did not show how we construct such self-adjoint extensions.
\end{enumerate}
\end{rem}
 Let us consider the equation

\begin{equation}
H\psi(x)=\pm i \lambda_0 \psi(x), \ \lambda_0 >0,\label{ma1}
\end{equation}
with  $H$ as in (\ref{eqz}). This equation known as Airy equation has two independents solutions $Ai(\cdot)$ and $Bi(\cdot)$  both in $L^2(-L,L)$(See 10.4.1 in \cite{Abr}). So the deficiency indices of $H$ are $(2,2)$ and we will show that the self-adjoint extensions are parameterized by a $U(2)$ matrices. By Theorem \ref{neu} we conclude that dimension of the factor space $\mathcal{E}=\mathcal{D}(H^*)/ \mathcal{D}(H)$ is $4$.\par
To study these self-adjoint extensions, we start by introducing the sesquilinear form, for $\phi,\psi \in \mathcal{D}_{max}$
$$\mathcal{B}(\phi,\psi)=\frac{1}{2i}(\langle H^*\phi,\psi \rangle - \langle \phi,H^* \psi \rangle).$$
$\mathcal{B}$  depends only on the boundary values of $\phi$ and $\psi$. When $\phi=\psi$ we get
$$\mathcal{B}(\phi,\phi)=\frac{1}{2i}(\phi'(L)\overline{\phi(L)}-\phi(L)\overline{\phi'(L)}-\phi'(-L)\overline{\phi(-L)}+\phi(-L)\overline{\phi'(-L)}).$$
Using parallelogram identity twice and the identities
$$\frac{1}{2i}(x\overline{y}-y\overline{x})=\frac{1}{4}(|x+iy|^2-|x-iy|^2);\ \forall x,y\in \mathbb{C},$$
and
$$ 2(x\overline{y}+y\overline{x})=|x+y|^2-|x-y|^2;\ \forall x,y\in \mathbb{C},$$ we get that:
\begin{eqnarray}\nonumber
4L\mathcal{B}(\phi,\phi) &=& \mid L\phi'(-L)-i\phi(-L)\mid^2+\mid L\phi'(L)+i\phi(L)\mid^2 \\
   &&-\mid L\phi'(-L)+i\phi(L)\mid^2-\mid L\phi'(L)-i\phi(L)\mid^2.\label{tah}
\end{eqnarray}
It is not obvious to conclude from the equation (\ref{tah}). As the factor space is of dimension $4$,
the boundary form $\mathcal{B}$ can be identified to the following skew linear form with $\mathbb{C}^4$ equipped with the standard hermitian metric.
$$\mathcal{B}:\mathbb{C}^4 \to \mathbb{C}$$
$$Z=(z_1,z_2,z_3,z_4)\mapsto \frac{1}{2i}(z_1\overline{z_2}-z_2\overline{z_1}-z_3\overline{z_4}+z_4\overline{z_3}).$$
This could be written as
$$\mathcal{B}(Z,Z)=\langle \left( \begin{array}{c}
z_1 \\
z_2 \\
z_3 \\
z_4
\end{array} \right), J\left( \begin{array}{c}
z_1 \\
z_2 \\
z_3 \\
z_4
\end{array} \right) \rangle.$$
With
$$J=\frac{1}{2}\quad
\begin{pmatrix}
0 & -i & 0&0 \\
i & 0& 0& 0\\
0&0&0&i\\
0&0&-i&0
\end{pmatrix}
\quad.$$
$$\mathcal{B}(Z,Z)=0\Leftrightarrow Z\perp J Z.$$
 We set
 $$P_+=\frac{1}{2}I+J,P_-=\frac{1}{2}I-J.$$  So we get the following properties
 $$J=\overline{J^t}, 4J^2=I.$$
 and
 $$P_+^2=P_+,P_-^2=P_-, P_+=P_+^*,P_-=P_-^*,$$
 $$P_+P_-=0, P_++P_-=I.$$
 So $P_+,P_-$ are orthogonal projectors. These matrices project the space
 $\mathbb{C}^4$ onto subspaces $\mathbb{C}^4_+=P_+\mathbb{C}^4$ and $\mathbb{C}^4_-=P_-\mathbb{C}^4$ and we get that
 $$\mathbb{C}^4=\mathbb{C}^4_+\oplus \mathbb{C}^4_-.$$
It turns out that  $J$-self-orthogonal subspaces of $\mathbb{C}^4$
 are in one to one correspondence with unitary operators acting from $\mathbb{C}^4_+$ onto $\mathbb{C}^4_-$. \newline Let $\mathcal{D}$ be a domain such that
$$\mathcal{D}(H)\subseteq \mathcal{D}\subseteq \mathcal{D}(H^*).$$
To any $\mathcal{D}$ corresponds an extension of the operator $H$.
$$H_{\mathcal{D}}\phi=H^*\phi,\ \forall \phi \in \mathcal{D}.$$
We denote by $\mathcal{D}^{\perp J}$ the space
$$\{x\in \mathcal{E}: B(x,y)=0,\ \forall y \in \mathcal{D}\}.$$ We have
$$ (H_{\mathcal{D}})^*=H_{\mathcal{D}^{\perp J}}.$$
The domain of a self-adjoint extension of $H$ is a subspace of $\mathcal{D}_{max}$, on which the sesquilinear form $\mathcal{B}(\phi,\phi)$ vanishes identically. So $H_{\mathcal{D}}$ is self-adjoint if only if
$$\mathcal{D}=\mathcal{D}^{\perp J}.$$
Below we show that these possible self-adjoint extensions of $H$ are parameterized by a unitary
matrix $U\in U(2)$.
\begin{lem}\label{jer}
\begin{enumerate}
\item Let $U$ be an unitary operator acting from $\mathbb{C}_+^4$ onto $\mathbb{C}^4_-$. Then the subspace $\mathcal{D}_U=\{v+Uv,v\in \mathbb{C}^4_+\}$ is $J$-self-orthogonal, that is
$$\mathcal{D}_U=\mathcal{D}_U^{\perp J}.$$
\item For very $J$-self-orthogonal subspace $\mathcal{D}$ of $\mathbb{C}^4$ there exists a unitary operator $U:\mathbb{C}^4_+\to \mathbb{C}^4_-$ such that
$$\mathcal{D}=\mathcal{D}_U.$$
\item The correspondence between $J$-self-orthogonal subspaces and unitary operators acting from $\mathbb{C}^4_+$ onto $\mathbb{C}^4_-$ is one to one;
$$U_1=U_2\Leftrightarrow \mathcal{D}_{U_1}=\mathcal{D}_{U_2}. $$
\end{enumerate}
\end{lem}
{\bf{The proof:}}
\begin{enumerate}
\item The mapping from $\mathbb{C}^4_+$ to $\mathcal{D}_U$ defined by $v\mapsto v+Uv$, is one to one. Indeed, the equality $v+Uv=0$ implies that $\parallel v\parallel=0$ as $v$ is orthogonal to $Uv \in \mathbb{C}^4_-$. So the mapping is bijective and we get
    $$dim (\mathcal{D}_U)=dim \mathbb({C}^4_+)=2.$$
    Let $v_1$ and $v_2$ be two arbitrary vectors of $\mathbb{C}^4_+$. We set $u_1=v_1+Uv_1$ and $u_2=v_2+Uv_2$. As $2J=P_+-P_-$ and $v_i=P_+v_i$ , $Uv_i=P_-Uv_i, i=1,2$ using the properties of $P_+$ and $P_-$, we get that $u_1$ and $u_2$ are $J$ orthogonal. So
    $$\mathcal{D}_U\subseteq (\mathcal{D}_U)^{\perp J}.$$ Since the Hermitian form $\mathcal{B}$, is non-degenerate on $\mathbb{C}^4$, then $\displaystyle dim (\mathcal{D}_U^{\perp J})=dim \mathbb{C}^4-dim (\mathcal{D}_U)=2.$ So
    $$\mathcal{D}_U= (\mathcal{D}_U)^{\perp J},$$
    i.e the subspace $\mathcal{D}_U$ is $J$-self-orthogonal.
    \item Let $\mathcal{D}$ be a $J$-self-orthogonal subspace. If
    $$v\in \mathcal{D}, v=v_1+v_2, v_1\in \mathbb{C}^4_+,v_2\in \mathbb{C}^4_-,$$
    the the condition of $J$ self-orthogonality; $v\perp_Jv$ means that $\langle v_1,v_1\rangle = \langle v_2,v_2\rangle.$ So, if $v_1=0$, the also $v$. This implies that the projection mapping $v\to P_+v$, considered as a mapping from $\mathcal{D}\to \mathbb{C}^4_+$ is injective. For a $J$-self-orthogonal subspace $\mathcal{D}$ of $\mathbb{C}^4$. The equality $dim (\mathcal{D})=dim (\mathbb{C}^4)-dim (\mathcal{D})$ holds.
    So $dim (\mathcal{D})=dim (\mathbb{C}^4_+)$ So, the injective linear mapping $v\to P_+v$ is surjective . The inverse mapping defined from $\mathbb{C}^4_-$ is presented in the form $v=v_1+Uv_1$, with $U$ is a unitary operator acting from $\mathbb{C}^+_4$ into $\mathbb{C}^4_-$. This mapping $v_1\to v_1+Uv_1$ maps the space $\mathbb{C}^4_+$ onto the subspace $\mathcal{D}$.\newline
    As $\langle v,Jv\rangle =0$ then $\langle v_1,v_1\rangle =\langle v_2,v_2\rangle , $ with $v_2=Uv_1$. As $v_1\in \mathbb{C}^4_+$ is arbitrary, this means that the operator $U$ is isometric. Since $dim \mathbb{C}^4_+=dim \mathbb{C}^4_-$ the operator is unitary. Thus, the originally given $J$-self-orthogonal subspace $\mathcal{D}$ is of the form $\mathcal{D}_U$. With $U$ is a unitary operator acting from $\mathbb{C}^4_+$ to $\mathbb{C}^4_-$.
    \item  The equality $\mathcal{D}_{U_1}=\mathcal{D}_{U_2}$ means that any vector of the form $v_1+Uv_1$ where $v_1\in \mathbb{C}^4_+$ can be represented in the form $v_2+Uv_2$ with some $v_2\in \mathbb{C}^4_+:$
        $$v_1+U_1v_1=v_2+U_2v_2.$$
        As $\displaystyle v_1,v_2\in \mathbb{C}^4-+, U_1v_1,U_2v_2\in \mathbb{C}^4_-$, then $v_1=v_2$ and $U_1v_1=U_2v_1$ for any $v_1\in \mathbb{C}^4_+$ which means that $U_1=U_2$. Thus
        $$\mathcal{D}_{U_1}=\mathcal{D}_{U_2}\Rightarrow U_1=U_2.$$
\end{enumerate}
This ends the proof of Lemma \ref{jer}.\hfill{$\square$}
Now we return to the equation (\ref{tah}) with boundary condition by setting  $z_1= L\phi'(-L)- i\phi(-L), z_2=L\phi'(L)+i\phi(L), z_3=L\phi'(-L)+i\phi(-L)$ and $z_4=  L\phi'(L)-i\phi(L)$ .Let us denote them by $H_U = (H, \mathcal{D}(U))$, here $\mathcal{D}$(U) is the space of functions
$\phi \in  \mathcal{D}_{max}$ satisfying by (\ref{tah}), the following boundary conditions
\begin{equation}\begin{pmatrix}
   L\phi'(-L)- i\phi(-L)\\
   L\phi'(L)+i\phi(L)
\end{pmatrix}=U \begin{pmatrix}
   L\phi'(-L)+i\phi(-L)\\
   L\phi'(L)-i\phi(L)
\end{pmatrix}.\label{equyahdi}
\end{equation}
These boundary conditions describe all the self-adjoint extensions $(H_U, \mathcal{D}(U))$ of $H$.
\begin{rem}
Let us point that the boundary condition (\ref{equyahdi}) is so important and could even break parity properties of solutions
of eigenfunctions equations and even in the case when the potential is of definite parity we can't say noting about the solutions.
\end{rem}
Using the fact that for $n$ order differential operator with deficiency indices $(n, n)$ all of its self-adjoint
extensions have a discrete spectrum, we conclude that all the spectra of the $H_U$ are totaly discrete. So the proof of Theorem \ref{th1} is ended. \hfill{$\square$}
For completeness let us recall  that a $2\times 2$ matrix $U$ with complexes coefficients is an element of $U(2)$ if and only if $U^*\cdot U=I_2$. So the determinant of $U$ is a complex of modulus $1$ and $det M: U(2)\to U(1)$ is a group homomorphism which is surjective and having the subgroup $SU(2)$ of matrices determinant one as a kernel. So
  $$U(1)\cong U(2)/SU(2).$$
By this we get the following parametrization of $U(2)$ and write that
\begin{equation}
U=e^{i\theta}M,\ det M=1, i.e \ M\in SU(2).
\end{equation}
For this let us recall some results and properties of $SU(2)$, representation.
\subsection{Representation and topology of $SU(2)$}
As we deal with matrices of order two  there is more explicit properties. Let $M\in SU(2)$ \begin{equation}
M=\begin{pmatrix}
   \alpha & \beta  \\
   \gamma & \lambda
\end{pmatrix}; M^*= \begin{pmatrix}
   \overline{\alpha} & \overline{\gamma}  \\
   \overline{\beta }& \overline{\lambda}
\end{pmatrix};
\end{equation}
using the fact $det(M)=\alpha \beta -\beta \gamma=1$, we get
\begin{equation} M^{-1}=\begin{pmatrix}
   \lambda & -\beta  \\
   -\gamma & \alpha
\end{pmatrix}.
\end{equation}
So
$$M^{-1}=M^{*}\Leftrightarrow \lambda=\overline{\alpha}; \ \rm{and}\ \ \gamma=-\overline{\beta},$$
and the generic form of matrices of $SU(2)$ is given by the following parametrization
\begin{equation}
M=\begin{pmatrix}
   \alpha & \beta  \\
   -\overline{\beta} & \overline{\alpha}
\end{pmatrix}
; |\alpha|^2+|\beta|^2=1.
\end{equation}
By taking $\displaystyle \alpha =\alpha_1+i\alpha_2$ and $\displaystyle \beta=\beta_1+i\beta_2,\alpha_i,\beta_i\in \mathbb{R}$, we get that
$$|\alpha_1|^2+|\alpha_2|^2+|\beta_1|^2+|\beta_2|^2=1.$$
This gives that $SU(2)$ as a topological space is holomorphic  to the sphere unity $S^3$ in $\mathbb{R}^4$. $SU(2)$ has three generators given Pauli matrices \cite{San}.
\begin{equation}
\tau_1=\begin{pmatrix}
   0 & 1  \\
   1& 0
\end{pmatrix}, \tau_2=\begin{pmatrix}
   0 & -i  \\
   i& 0
\end{pmatrix}, \tau_3=\begin{pmatrix}
   1 & 0  \\
   0& -1
\end{pmatrix}.
\end{equation}
We write
$$M=\alpha_1I_2-i(\alpha_2,\beta_1,\beta_2).(\tau_1,\tau_2,\tau_3).$$
\subsection{Form of solutions}
\begin{figure}[htp]
\centering
{\includegraphics[width=0.45\textwidth,height=0.35\textwidth,angle=0]{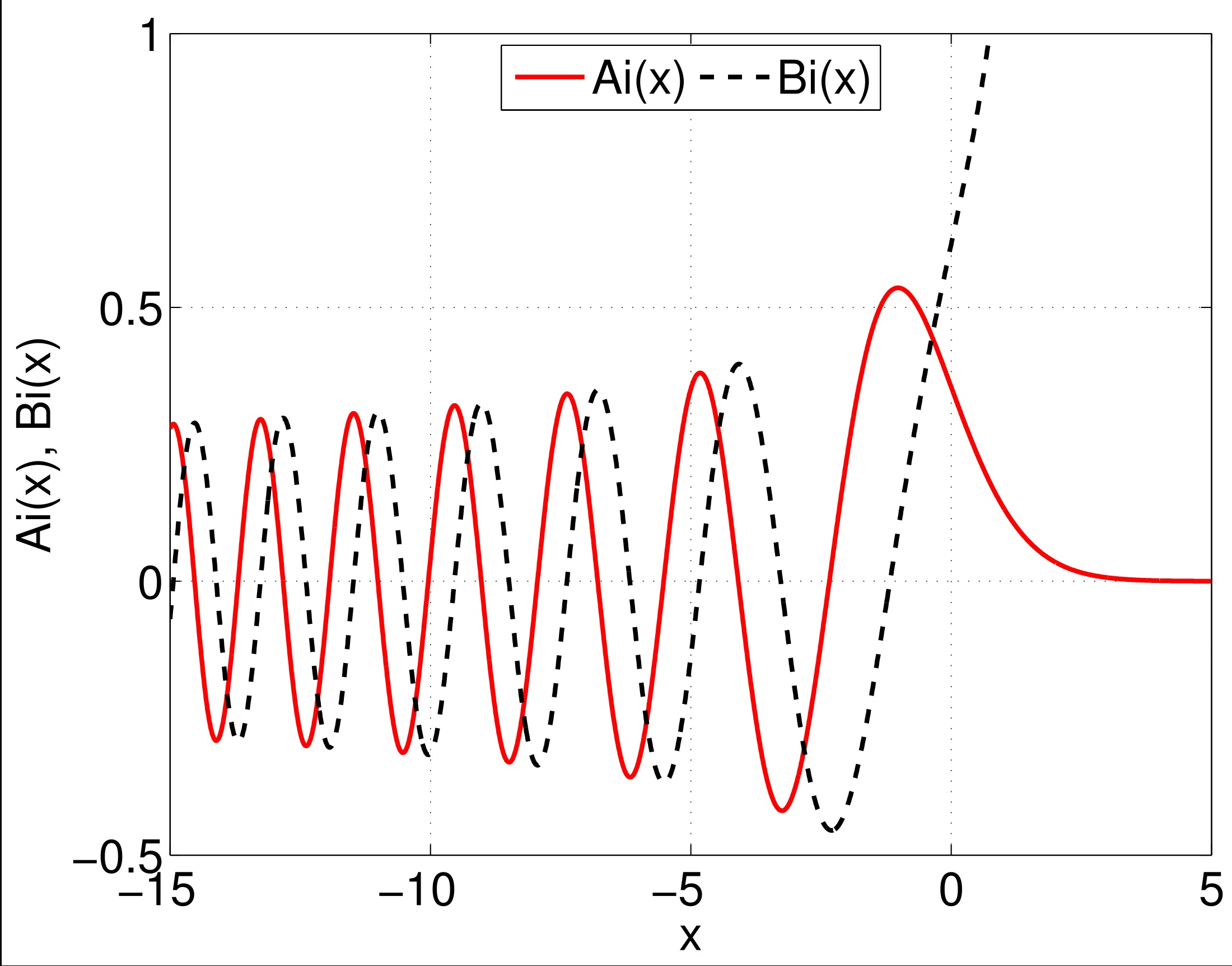}}
{\includegraphics[width=0.45\textwidth,height=0.35\textwidth,angle=0]{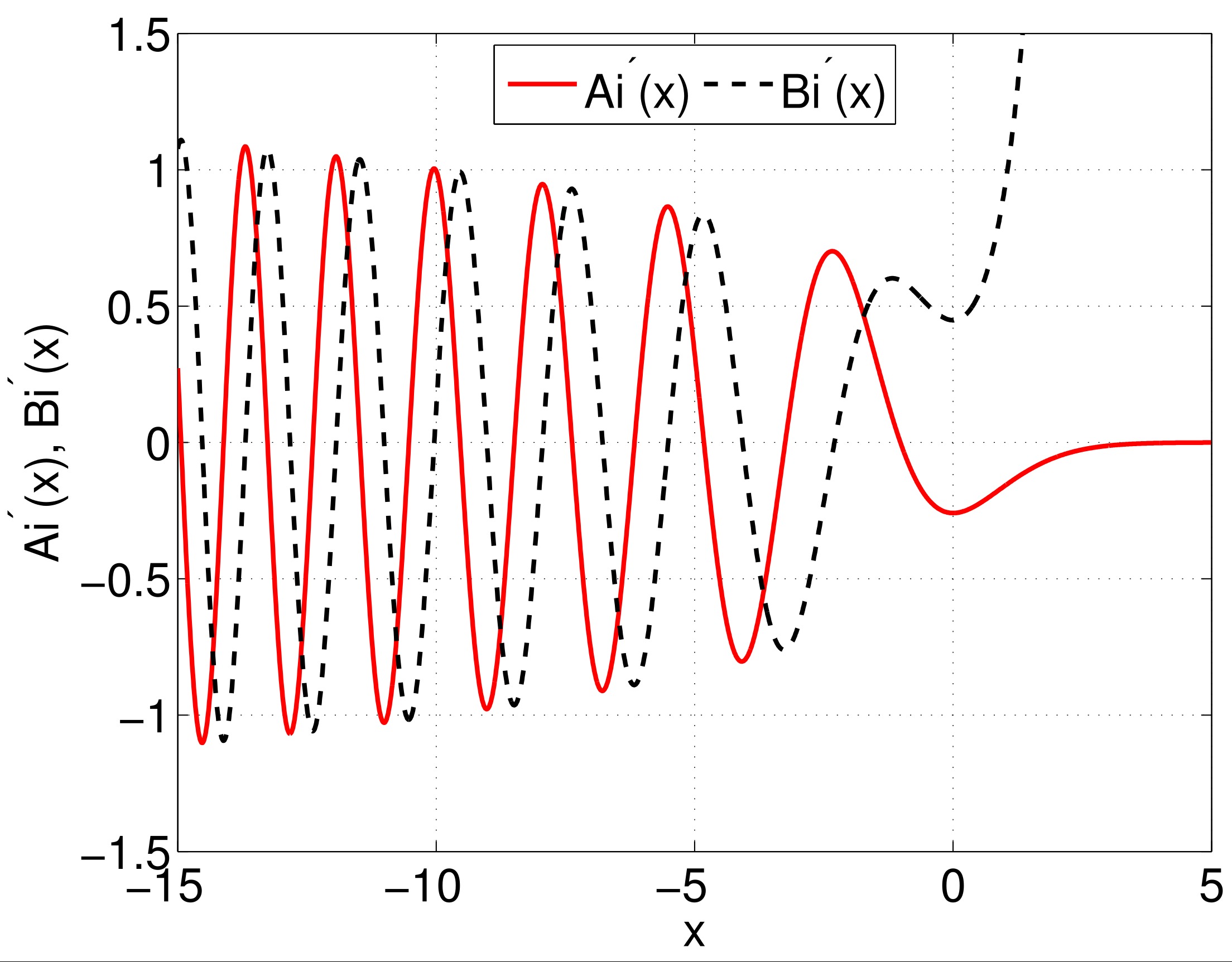}}
\caption{Airy functions and the corresponding derivatives.}\label{airy}
\end{figure}

The spectral equation associated to stark operator has been solved by Airy special functions
$\displaystyle Ai(\cdot)$ and $\displaystyle Bi(\cdot)$ \cite{far, val1}, see Figures \ref{airy}, which are the solution of the following second order differential equation
\begin{equation}-\frac{d^2 \psi}{dx^2}(x)+Fx\psi(x)=E\psi(x). \label{k1}
\end{equation}
Using the change of variable:
$$\xi=\frac{E}{F\rho}; \rho=F^{-\frac{1}{3}},\ x=\rho z,$$
we get the new equation
\begin{equation}
\psi''(z)=(z-\xi)\psi(z).\label{t1}
\end{equation}
The solutions of equ. (\ref{k1}) are two linearly independent
Airy functions $\displaystyle Ai(z-\xi)$ and $\displaystyle Bi(z-\xi)$. The eigenfunctions associated to the equation (\ref{t1}) are given as a superposition of two linearly independent functions of the form
\begin{equation}
\phi(z)= A\cdot Ai(z-\xi)+B\cdot Bi(z-\xi); \ \pmb{\Phi}= \begin{pmatrix}
   A\\
   B
\end{pmatrix}\in \mathbb{R}^2.\label{zoh}
\end{equation}
\begin{rem}
At this stage, lets remark that works dealing with a half line domaine, i.e, with a potential $V(x)=0$ for $x\leq0$ and $V(x)=+\infty$ for $x>0$; in (\ref{zoh}) we get just $Ai(\cdot)$ and the quantized energies are then given in terms of the
zeros of the well-behaved Airy  $Ai(\cdot)$. So the eigenvalues of the operator are given by $E=F^{\frac{2}{3}}\xi$, with $-\xi$ are the $k$-th zero of $Ai$.
\end{rem}
The solutions of equation (\ref{t1}) are of the form
\begin{equation}
\phi(x)=A\cdot Ai(F^{\frac{1}{3}}(x-\frac{E}{F}))+B\cdot Bi(F^{\frac{1}{3}}(x-\frac{E}{F})).
\end{equation}
We set
\begin{equation}
L^+(E,F)=F^{\frac{1}{3}}(L-\frac{E}{F}),\label{la1}
\end{equation}
and
\begin{equation}
L^-(E,F)=-F^{\frac{1}{3}}(L+\frac{E}{F}).\label{la2}
\end{equation}
So
\begin{eqnarray*}
&& \begin{pmatrix}
   L\phi'(-L)- i\phi(-L)\\
   L\phi'(L)+i\phi(L)
\end{pmatrix}  \\
   &=& \begin{pmatrix}
   L (A\cdot Ai'(L^-(E,F))+B\cdot Bi'(L^-(E,F)))-i(A\cdot Ai(L^-(E,F))+B\cdot Bi(L^-(E,F)))\\
  L (A\cdot Ai'(L^+(E,F))+B\cdot Bi'(L^+(E,F)))+i(A\cdot Ai(L^+(E,F))+B\cdot Bi(L^+(E,F)))
\end{pmatrix}  \\
   &=& \begin{pmatrix}
   A(L\cdot Ai'(L^-(E,F))-iAi(L^-(E,F)))+B(L\cdot Bi'(L^-(E,F))-i Bi(L^-(E,F)))\\
  A (L\cdot Ai'(L^+(E,F))+i Ai(L^+(E,F)))+B(L\cdot Bi'(L^+(E,F))+iB\cdot Bi(L^+(E,F)))
\end{pmatrix} \\
&=&\mathcal{L}(\xi)\pmb{\Phi}.
\end{eqnarray*}
With $$ \mathcal{L}(\xi)=\begin{pmatrix}
   L\cdot Ai'(L^-(E,F))-iAi(L^-(E,F)) & L\cdot Bi'(L^-(E,F))-i Bi(L^-(E,F))\\
  L\cdot Ai'(L^+(E,F))+iAi(L^+(E,F)) &L\cdot Bi'(L^+(E,F))+i Bi(L^+(E,F))
\end{pmatrix}.
$$
and
\begin{eqnarray*}
   &&\begin{pmatrix}
   L\phi'(-L)+i\phi(-L)\\
   L\phi'(L)-i\phi(L)
\end{pmatrix} \\
   &=&  \begin{pmatrix}
   A(L\cdot Ai'(L^-(E,F))+i Ai(L^-(E,F)))+B(L\cdot Bi'(L^-(E,F))+i Bi(L^-(E,F)))\\
  A (L\cdot Ai'(L^+(E,F))-i Ai(L^+(E,F)))+B(L\cdot Bi'(L^+(E,F))-iB\cdot Bi(L^+(E,F)))
\end{pmatrix} \\
   &=&  \mathcal{M}(\xi)\pmb{\Phi}.
\end{eqnarray*}
With $$ \mathcal{M}(\xi)=\begin{pmatrix}
   L\cdot Ai'(L^-(E,F))+iAi(L^-(E,F)) & L\cdot Bi'(L^-(E,F))+i Bi(L^-(E,F))\\
  L\cdot Ai'(L^+(E,F))-iAi(L^+(E,F)) &L\cdot Bi'(L^+(E,F))-i Bi(L^+(E,F))
\end{pmatrix}.
$$
Using (\ref{equyahdi}) we get the following relation between $\mathcal{L}(\xi)$ and $\mathcal{M}(\xi)$.
\begin{equation}
(\mathcal{L}(\xi)-U\mathcal{M}(\xi))\pmb{\Phi}=0.
\end{equation}
To get a nontrivial solution to (\ref{zoh}), we need that $(\mathcal{L}(\xi)-U\mathcal{M}(\xi))$ be not invertible which is equivalent to
\begin{equation}
det(\mathcal{L}(\xi)-U\mathcal{M}(\xi))=0.\label{zoh2}
\end{equation}
Unfortunately it is not possible to get a simple analytic expression for the equation (\ref{zoh2}). Below, we give some particular cases which allow us to simplify  least a little bit the general expression.
\section{Interesting particular cases}
In this section, we consider four particular cases of $U$. They are the most interesting and generally studied in literature \cite{naj1,olen, ReSi4,val1}, known as Dirichlet, Neumann, Dirichlet-Neumann conditions and others. In general, it is not trivial to solve explicitly the determinant equations (\ref{zoh2}). In \cite{far}, the authors used numerical methods. Namely, the classical "Newton method" in "Mathematica tools" by "Find Root". Here, we implement a combination of the Bisection and the Newton methods. We approximate the zeros of the determinants with a maximal error $10^{-8}$. Below, we consider some particular cases, which allow us to perform interesting computational results. For a fixed interval length $L$, we compute the first four eigenvalues for different fields $F$. Thereafter, for fixed $F$, we determined the first four eigenvalues for different width $L$ of the quantum well. The associated eigenfunctions are also plotted.
\begin{enumerate}
\item The case $U=I$.\\  This case leads to the operator $H_I=(H,\mathcal{D}(I))$ known as Dirichlet operator, with
\begin{equation}
\{\phi \in L^{2}([-L,L]), H_{I_2}\phi\in L^{2}([-L,L]) \ {\rm{and}}\ \phi(-L)=\phi(L)=0\}. \label{zokzok}
\end{equation}
So
$$\mathcal{L}(\xi)-U\mathcal{M}(\xi)=\mathcal{L}(\xi)-\mathcal{M}(\xi)=2i\begin{pmatrix}
-Ai(L^-(E,F)) &  -Bi(L^-(E,F))\\
  Ai(L^+(E,F)) & Bi(L^+(E,F))
\end{pmatrix}.$$
The equation (\ref{zoh2}) yields to
\begin{equation}Ai(L^-(E,F))Bi(L^+(E,F))-Ai(L^+(E,F))Bi(L^-(E,F))=0.\label{k2}
\end{equation}
To get the representation of the eigenfunction $\phi_n(x)$ associated to the eigenvalue $E_n$ already calculated and given in table 1. We use the equation (\ref{zoh}) and the boundary conditions given in (\ref{zokzok}) to obtain
\begin{equation}
A\cdot Ai(L^+(E,F))+B\cdot Bi(L^+(E,F))=0,
\end{equation}
and
\begin{equation}
A\cdot Ai(L^-(E,F))+B\cdot Bi(L^-(E,F))=0.
\end{equation}
This gives that that $$A=-B\frac{Bi(L^-(E,F))}{Ai(L^-(E,F))}=-B\frac{Bi(L^+(E,F))}{Ai(L^+(E,F))}.$$
So finally, we get that
\begin{equation}
\phi_n(x)=C\big[Bi(L^-(E,F))\cdot Ai(F^{\frac{1}{3}}(x-\frac{E_n}{F}))-Ai(L^-(E,F))\cdot Bi(F^{\frac{1}{3}}(x-\frac{E_n}{F}))\big],
\end{equation}
with $C\in\mathbb{R}$ and
\begin{equation}
L^+(E,F)=F^{\frac{1}{3}}(L-\frac{E}{F})\qquad\text{and} \qquad
L^-(E,F)=-F^{\frac{1}{3}}(L+\frac{E}{F}).\label{la12}
\end{equation}
In table 1, we give the eigenvalues for different cases. It should be stressed that an interesting effect appears by varying $L$ and $F$.
\begin{table}[htp]\label{tab1}
\[\begin{array}{|ll|llll|}
 \hline 					
L	&F	    & E_1	    &E_2	    &E_3	    &E_4\\
 \hline
1   & 0 &  2.4674 &9.8696&22.2066&39.4784\\
1	&0.01	& 2.4673	&9.8696	&22.2066	&39.4784\\
1	&0.1	& 2.4672	&9.86965&22.2066	&39.4784\\
1	&1	    & 2.4498	&9.8748	&22.2097	&39.4803\\
1	&5	    & 2.0416	&9.9877	&22.2841	&39.5261\\
\hline
\hline 		
1	&1	&2.4498	&9.8748	&22.2097&39.4803 \\
2	&1	&0.3554	&2.5324	&5.6007	&9.9001 \\
3	&1	&-0.6618&1.0947	&2.6628	&4.5376	 \\
4	&1	&-1.6618&0.0879 &1.5216	&2.8152		 \\
\hline
\end{array} \]
  \caption{Eigenvalues of the case 1}
\end{table}

\item The case $U=-I$. \\ This particular case leads to the operator $H_{-I}=(H,\mathcal{D}(-I))$ known as Neumann operator, with
\begin{equation}
\{\phi \in L^{2}([-L,L]), H_U\in L^{2}([-L,L]) \ {\rm{and}}\ \phi'(-L)=\phi'(L)=0\}.
\end{equation}
So
$$\mathcal{L}(\xi)-U\mathcal{M}(\xi)=\mathcal{L}(\xi)+\mathcal{M}(\xi)=2L\begin{pmatrix}
Ai'(L^-(E,F)) & Bi'(L^-(E,F))\\
  Ai'(L^+(E,F)) & Bi'(L^+(E,F))
\end{pmatrix}.$$
The equation (\ref{zoh2}) yields to
\begin{equation}Ai'(L^-(E,F))Bi'(L^+(E,F))-Ai'(L^+(E,F))Bi'(L^-(E,F))=0.\label{om1}
\end{equation}
For the eigenfunctions we get:
\begin{equation}
\phi_n(x)=C\big[Bi'(L^-(E,F))\cdot Ai(F^{\frac{1}{3}}(x-\frac{E_n}{F}))-Ai'(L^-(E,F))\cdot Bi(F^{\frac{1}{3}}(x-\frac{E_n}{F}))\big],
\end{equation}
\begin{table}[htp]\label{tab2}
\[\begin{array}{|ll|llll|}
 \hline 					
L	&F	    & E_1	    &E_2	    &E_3	    &E_4\\
 \hline
1&0& 0& 2.4674&9.8696& 22.2066\\
1&	0.01& -0.00001 & 2.4674& 9.8696&22.2066	\\
1&	0.1	& -0.0013 & 2.4684& 9.8697&	22.2066	\\
1&	1   & -0.1278& 2.5674& 9.8825&	22.2125	\\
1&	5   & -2.0330& 3.7841& 10.215&	22.3241	\\
\hline
 \hline 	
1&1   & -0.1278& 2.5674& 9.8825&22.2112	\\
2&1	  &-0.9818 &1.1254	&2.7014	&5.6284	\\
3&1	  &-1.9812 &0.2475	&1.7735	&2.9509	 \\
4&1	  &-2.9812& -0.7518 &0.8199	&2.1551	 \\
\hline
\end{array} \]
  \caption{Eigenvalues of the case 2}
\end{table}	

\item The case $U=\begin{pmatrix}
1 & 0\\
  0& -1
\end{pmatrix}.$ \begin{equation}
\{\phi \in L^{2}([-L,L]), H_U\in L^{2}([-L,L]) \ {\rm{and}}\ \phi(-L)=\phi'(L)=0\}.
\end{equation} In this particular case we get $$\mathcal{L}(\xi)-U\mathcal{M}(\xi)=2\begin{pmatrix}
-iAi(L^-(E,F)) &  -iBi(L^-(E,F))\\
  LAi'(L^+(E,F)) & LBi'(L^+(E,F))
\end{pmatrix}.$$
The equation (\ref{zoh2}) yields to
\begin{equation}Ai'(L^+(E,F))Bi(L^-(E,F))-Ai(L^-(E,F))Bi'(L^+(E,F))=0.
\end{equation}
For the eigenfunctions we get that
\begin{equation}
\phi_n(x)=C\big[Bi(L^-(E,F))\cdot Ai(F^{\frac{1}{3}}(x-\frac{E_n}{F}))-Ai(L^-(E,F))\cdot Bi(F^{\frac{1}{3}}(x-\frac{E_n}{F}))\big],
\end{equation}
\end{enumerate}						
\begin{table}[htp]\label{tab3}
\[\begin{array}{|ll|llll|}
 \hline 				
L	&F	    & E_1	    &E_2	    &E_3	    &E_4\\
 \hline
1&0&0.6168&5.5516&15.4212&30.2256\\
1&	0.01& 0.6208&	5.5521	&15.4214&	30.2257\\
1&	0.1	&  0.6570&	5.5563	&15.4229&	30.2265 \\
1&	1	  &0.9864	&5.6153	&15.4432	&30.2367\\
1&	5	  &1.6096	&6.3689	&15.6591	&30.3396 \\
\hline
\hline 					
1&	1&	0.9864&	5.6153&	15.4432&	30.2367 \\
2&	1&	0.3175&	1.8336&	3.9959	&7.6204 \\
3&	1&-0.6619&	1.0798&	2.3777&	3.6819	 \\
4&	1&-1.6618&	0.0879&	1.5192&	2.7497	 \\
\hline
\end{array} \]
  \caption{Eigenvalues of the case 3}
\end{table}
\begin{rem}
The absence of splitting and the shift phenomena in the non-degenerates case found in the previous three cases corresponds to the vanishing of the linear stark effect in the perturbation theory.
\end{rem}
\begin{figure}[htp]
\centering
\subfloat[Case 1]
{\includegraphics[width=0.325\textwidth,height=0.25\textwidth,angle=0]{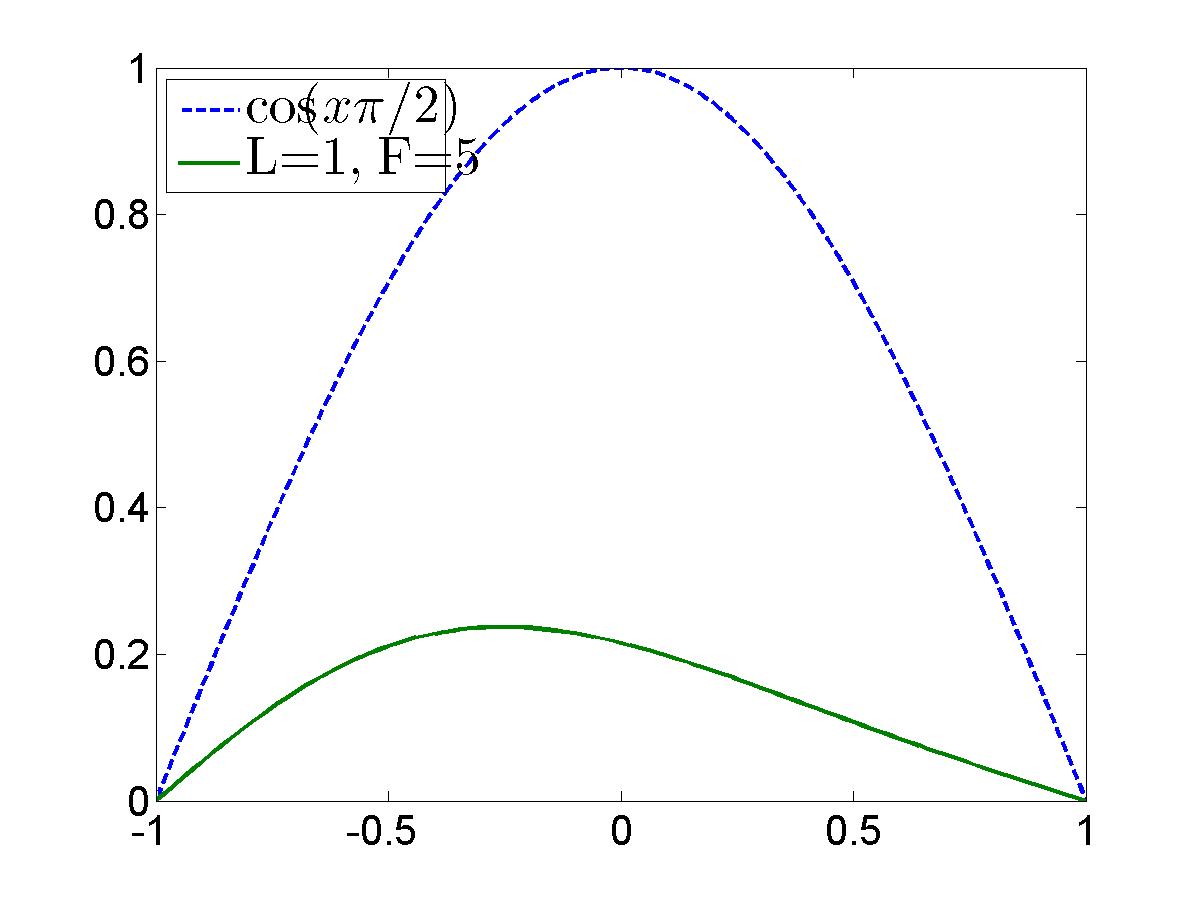}}
\subfloat[Case 2]
{\includegraphics[width=0.325\textwidth,height=0.25\textwidth,angle=0]{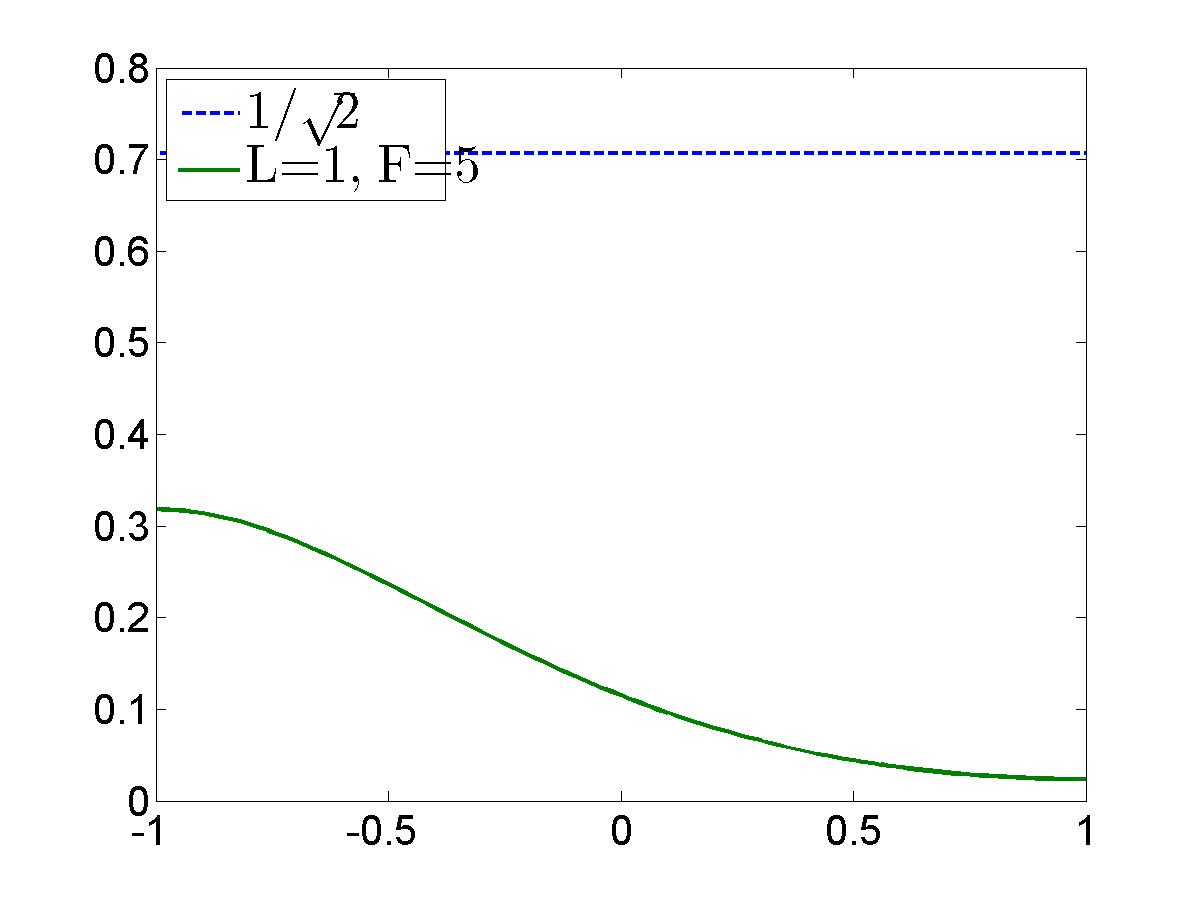}}
\subfloat[Case 3]
{\includegraphics[width=0.325\textwidth,height=0.25\textwidth,angle=0]{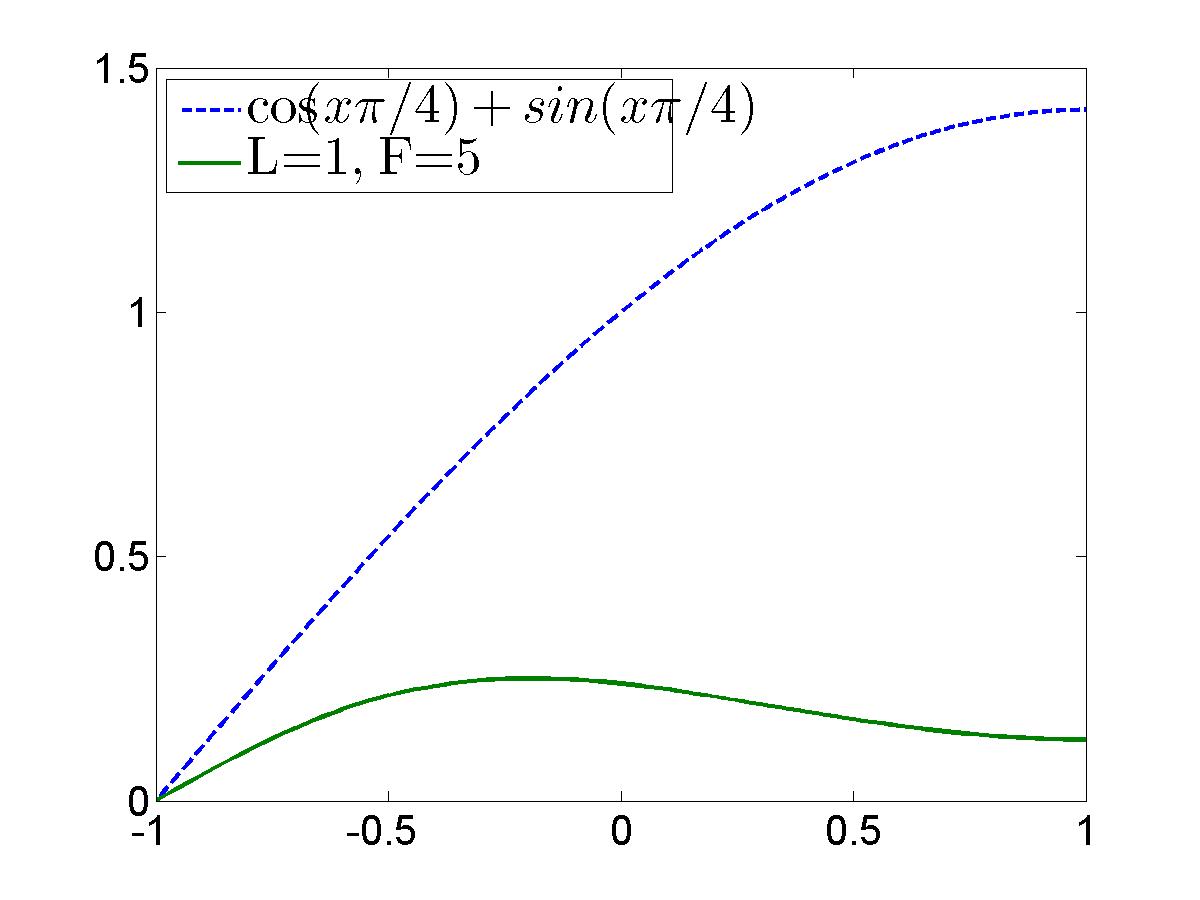}}
\caption{Comparing the analytical eigenfunction for $L=1$, $F=0$ to the computational result for $L=1$, $F=5$.}\label{fig9}
\end{figure}
\begin{rem}It is important to note that, for the three previous cases, the eigenvalues decreases when $L$ increases. This behavior is similar to the free case. See tables 1, 2 and 3. In figure \ref{fig9}, we remark that in the three cases, we have concentration of the eigenfunction on the left of the well when $F\neq 0$. i.e the particle is shifted to the left to minimize the total energy.
\end{rem}
\newpage
\subsection{Splitting phenomena}
\label{split}
\begin{figure}[htp]
\centering
{\includegraphics[width=0.65\textwidth,height=0.35\textwidth,angle=0]{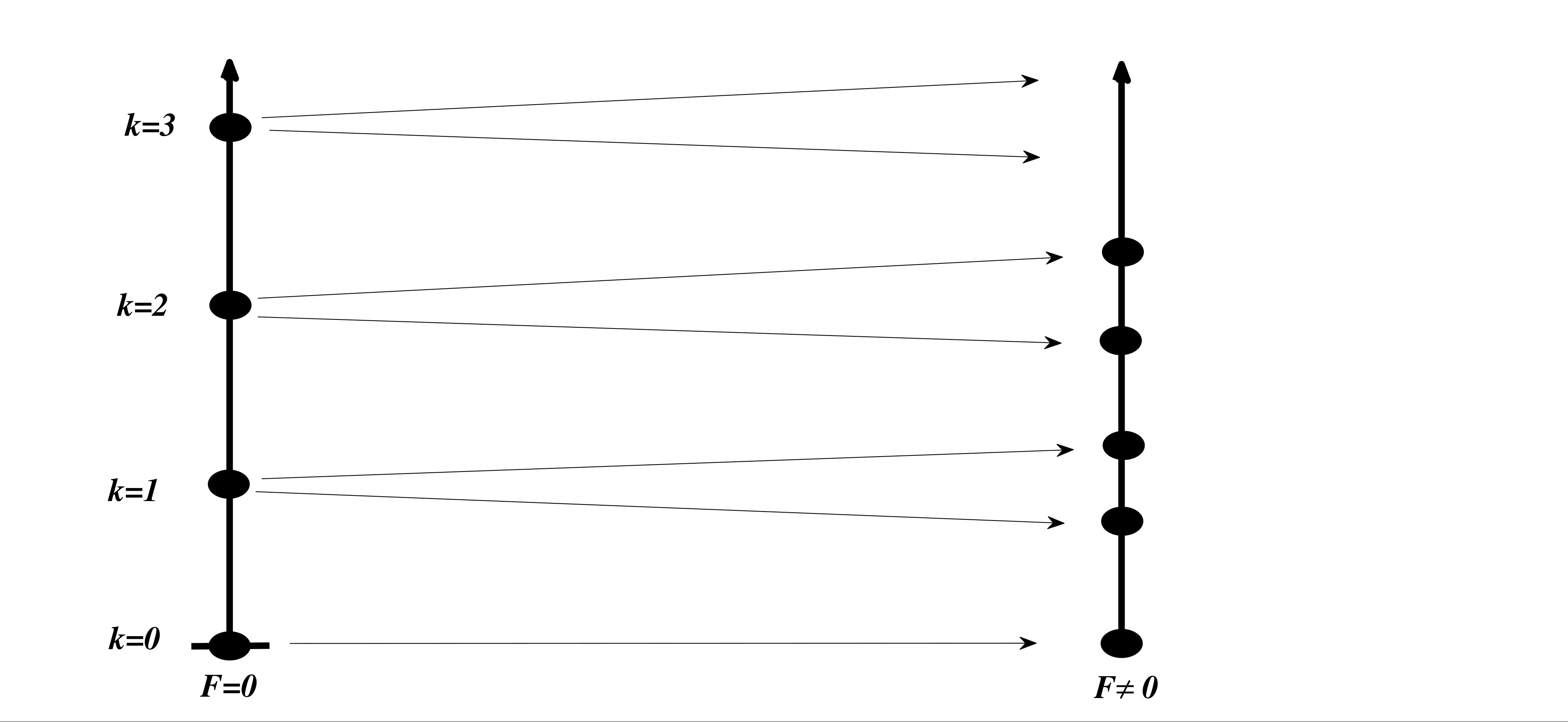}}
\caption{Splitting: Compared with other three cases. It is a non-degenerates eigenvalues and the stark effect was a shift of eigenvalues. In the current case it is degenerate and splitting.} \label{splits}
\end{figure}

In this subsection, we will consider the case where $U=\tau_1=\begin{pmatrix}
0 & 1\\
  1& 0
\end{pmatrix}.$
It leads to the operator $H_{\tau_1}=(H,\mathcal{D}(\tau_1))$, with
\begin{equation}
\mathcal{D}(\tau_1)=\{\phi \in L^{2}([-L,L]), H_U\in L^{2}([-L,L]) \ {\rm{and}}\ \phi'(-L)=\phi'(L), \phi(L)=\phi(-L)\}.
\end{equation}
This case is not considered in literature. We shad some lights on the spectral theory on $H_U$. We expect it modeled the system which highlights the phenomena of splitting that is long sought by physician. Let's recall that since 1913, J. Stark stated that, when a particle is exited a strong electric field splits on number of components an effect that goes after his name. The observed splitting agree with the calculation developed in this work. Which confirm the accuracy of the implemented numerical methods used here. The splitting is symmetrical in the where the field $F=0$, see Figure \ref{splits}.
\par Mathematically there is a deep relation between degeneracy and symmetry. This implies the existence of conjugation under which the operator remains unchanged. Such question is related to the theory of the symmetry group of the operator. The possible degeneracies of the eigenvalues with a particular symmetry group of the operator is specified by dimensionality of the irreducible representation of the group.
The eigenfunction corresponding to $m$-degenerates eigenvalues form a basis for a $m$-dimensional irreducible representation of the symmetry group of the operator.
\par The degeneracy could arises due to the presence of some kind of symmetry in the system under consideration or related a characteristic of dynamical symmetry of the system. It also could be connected to the existence of bound orbits in the classical physics. The degeneracy in the present case is abolished when the symmetry is bracken by the presence of external electric field $F$. This engender the splitting in the degenerate energy level accrurating the numerical part of the proved result. We notes that the first order Stark effect is zero for the ground state (like Hydrogen atom).
\par The equation (\ref{zoh2}) yields to
\begin{multline*}
[(Ai'(L^-(E,F))-Ai'(L^+(E,F))(Bi(L^+(E,F))-Bi(L^-(E,F))]\\ -[(Ai(L^+(E,F))-A(L^-(E,F))(Bi'(L^-(E,F))-Bi'(L^+(E,F))]=0.
\end{multline*}
For the eigenfunctions, we get that
\begin{eqnarray}\nonumber
\phi_n(x)&=& C\big[(Bi'(L^+(E,F))-Bi'(L^-(E,F)))\cdot Ai(F^{\frac{1}{3}}(x-\frac{E_n}{F}))\\
&&+(Ai'(L^-(E,F))-Ai'(L^+(E,F)))\cdot Bi(F^{\frac{1}{3}}(x-\frac{E_n}{F}))\big],
\end{eqnarray}
\begin{table}[htp]
\[\begin{array}{|ll|lllll|}
 \hline 				
L	&F	   & E_0  & E_1	    &E_2	    &E_3	    &E_4\\
\hline 	
1&0&  0.0& 9.8696&39.4784&--&--\\				
1&	0.01   &0.0    &9.86796  & 9.87119	 &39.4778  &39.47891\\
2&	0.01   &0.0    &2.46422  & 2.47705   &9.86800  &9.87119\\
3&	0.01   &0.0    &1.09189  & 1.10144   &4.34411  &4.38889\\
4&	0.01   &0.0    &0.61063  & 0.62336   &2.46426  &2.47063 \\
\hline
\end{array} \]
  \caption{Case 4}
\end{table}
\begin{rem}
For the splitting case, we get the non-zero case between two eigenvalues of the Stark operator, except for the ground state, see table 4. Moreover, to get a significant figure satisfying the boundary conditions, we used a numerical precision up to $10^{-8}$. Indeed, if we used a less precisions some of the geometrical behaviors of eigenfunctions are in general not representative, see figure \ref{splits}.
\end{rem}
\newpage
\section{Concluding remarks}
In this work, we presented an analytical and computational study of Stark operators and precisely the self-adjoint operators on finite domains.
We numerically analyzed interesting boundary conditions. Even, we used a lot of approximations, the presented computational result confirm and accurate all analytical ones. The splitting phenomena developed in this work indicates a perfect start to develop other similar physical result, namely in the proper Stark effect. We intend to study more realistic model by considering random behavior of electric fields.
\subsection*{Acknowledgements}
We wish to thank  Professor C. Beddani for providing us interesting references and professor J. Faraut for valuable comments.


\begin{thebibliography}{99}
\smallskip
\bibitem{Abr} Abramowitz M.; Stegun, I. A.: {\sl\textit{ Mathematical Functions with Formulas, Graphs, and Mathematical Tables}}, 9th printing. New York: Dover, pp. 446-452, 1972.
 \bibitem{Ak}   Akhiezer N. I., Glazman, I. M.: {\sl\textit{Theory of linear operators in Hilbert space, Frederick
Ungar Publishing Company}} New-York (1961).
 \bibitem{bal}  Balasubramanian V., S. Das, E. C. Vagena: {\sl\textit{Generalized uncertainty principle and self-adjoint operators}} Ann. of Phy. 360 p 1-18 (2015).
 \bibitem{ver}  De Verdi\`ere Y. C. {\sl\textit{ Confining quantum particles with a purely magnetic field.}} Annales de l'institut Fourier, 60 no. 7, p 2333-2356 (2010).
    \bibitem{Bon} Bonneau G,  Faraut J.   Valent G,: {\sl\textit{Self-adjoint extensions of operators and the teaching of quantum mechanics}}
American Jour.  Phy.  69, 322 (2001); doi: http://dx.doi.org/10.1119/1.1328351
\bibitem{dev} Devinatz A.:{\sl\textit{The deficiency index of ordinary self-adjoint differential
operators.}} Paci. Jour. Math. 16 no 2  p 243-262 (1966).
\bibitem{Dom} Domingos J. M., Caldeira M. H.: {\sl\textit{Self-adjointness of momentum operators in generalized coordinates.}} Foundations of Physics February 1984, Volume 14, Issue 2, pp 147-154.
\bibitem{grb} Grubb G.: {\sl\textit{A characterization of the non-local boundary value problems associated with an elliptic operator}} Ann. Scula Norm. Sup. Pisa 22, p 425-513, (1968).
\bibitem{Eva} Evans W. D., Ibrahim, S. E.: {\sl\textit{Boundary conditions for general ordinary differential operators and their adjoints}} Proc. Royal Soc. Edinburgh 114 A p99-117 (1990).
\bibitem{Eve} Everitt W. N.: {\sl\textit{Integrable-square solutions of ordinary differential equations.}} (III) Quart. J. of Math. (Oxford) 14 (2)p 170-180 (1963).
     \bibitem{Eve2} Everitt  M. L.: {\sl\textit{Complex symplectic geometry with applications to ordinary differential operators}} Trans. Amer. Math. Soc. 351 p 4905-4945 (1999).
\bibitem{fac} Facchi P., G. Garnero, M. Ligabo: {\sl\textit{Self-adjoint extensions and unitary operators on the boundary}} March 14 (2017) arXiv: 1703.04091v1.
\bibitem{far}  Farhang M. L.; Hasan Bouzari, H., Ahmadi, F.: {\sl\textit{Solving Schrodinger equation specializing to the Stark effect
in linear potential by the canonical function method}} Jour. Theo. and App. Phy (2014)  8:140
DOI 10.1007/s40094-014-0140-x.
\bibitem{Fu}  Fu S. Z.: {\sl\textit{On the Self-Adjoint Extensions of Symmetric Ordinary Differential Operators in Direct Sum Spaces}} Jour. Diff. Equ. (100) p 269-291. (1992).
\bibitem{Hip}  Hiptmair R., Kotiuga, P. R.; Sebastien Tordeux, S.:{\sl\textit{Self-adjoint curl operators}} Annali di Matematica Pura ed Applicata, Springer Verlag,  191 (3), p.431-457 (2012)
  \bibitem{ibo} Ibort A.: {\sl\textit{ Three lectures on global boundary conditions and the theory of selfadjoint extensions of the covariant Laplace-Beltrami and Dirac operators on Riemannian manifolds with boundary}}  Arxiv: 1205.3579
\bibitem{Kat}   Katsnelson  V.:  {\sl\textit{ Self-adjoint boundary conditions for the prolate spheroid differential operator}} https://arxiv.org/pdf/1603.07542.pdf
      \bibitem{Mil} Miller D. A. B.; Chemla, D. S.; Damen, T. C.; Gossard, A. C.; Wiegmann, W; Wood, T. H.;
Burrus, C. A.: {\sl\textit{ Electric field dependence of optical absorption near the bandgap of
quantum well structures.}} Phys. Rev. B Vol. 32 1043.
\bibitem{Nai} Naimark M. A.:  {\sl\textit{Linear differential operators.}} vol 2, Frederick Ungar Publishing Company,
New-York (1968).
\bibitem{naj1} Najar H., Raissi M.: {\sl\textit{Eingenvalue Asymptotics for the Stark operator}}, in progress.
\bibitem{naj2} Najar H., Zahri M.: {\sl\textit{Domain dependent Random Stark operators on a quantum Well}}, In progress.
\bibitem{von} Von Neumann J.:  {\sl\textit{Allgemeine Eigenwerthorie Hermitescher Funktionaloperatoren.}} Math. Ann. 102 p 49-131 (1930).
\bibitem{olen} Olendski O.:  {\sl\textit{ Comparative analysis of electric field influence on the quantum wells with different boundary
conditions}} Annalen Phys. 527, p 278-295 (2015).
\bibitem{oli} De Olivera C. R.: {\sl\textit{Intermidate spectral theory and quantum dynamics}} Progress in math. phy. Vol54 Birkh\"a sser, Belin (2009).
\bibitem{kal} Kalf H., Schminke U.V., Walter, J., Wust, R.:  {\sl\textit{ On the spectral theory of
Schr\"odinger and Dirac operators with strongly singular potentials}} in Lecture
Notes in Mathematics 448 , p 182-226 (1975).
\bibitem{Rob} Robinett R. W.: {\sl\textit{ The Stark effect in linear potentials.}}
 European J. Phys. 31 , no. 1, p 1-13 (2010).
\bibitem{ReSi}  Reed M., Simon, B.: {\sl\textit{Methods of modern mathematical physics. III. Scattering theory.}} Academic Press [Harcourt Brace Jovanovich, Publishers], New York-London, (1979).
\bibitem{ReSi4}  Reed M., Simon, B.: {\sl\textit{Methods of modern mathematical physics. IV. Analysis  of Operators.}} Academic Press [Harcourt Brace Jovanovich, Publishers], New York-London, 1977.
\bibitem{sim}  Simon B.: {\sl\textit{Essential selfajointeness of Schr\"odinger operators with singular potentials}} Archive for Rational Mechanics and Analysis , {\bf{ 52}}, Issue 1, p 44-48 ( 1973).
\bibitem{San} Santilli R. M.: {\sl\textit{ Isorepresentations of the Lie-isotopic SU(2) algebra with applications to nuclear physics and to local realism. }} Acta Appl. Math. 50 , no. 1-2, p 177-190 (1998).
\bibitem{Sun1} Sun J.: {\sl\textit{ On the self-adjoint extensions of symmetric spaces  ordinary differential operators with middle deficiency indices}} Acta Math. Sincia. New series 2 (2) p152-167. (1986).
   \bibitem{von} von Neumann J.: {\sl\textit{Allgemeine Eigenwerttheorie hermitescher Funktionaloperatoren.}} Math. Ann. 102, p 49-131, (1929).
 \bibitem{val1} Vallee O., Soares M.: {\sl\textit{Airy Functions and Applications to Physics.}}
Imperial College Press.
\bibitem{Wie} Wiegmann, W, Wood T. H.; Burrus, C. A.: {\sl\textit{ Electric field dependence of optical absorption near the bandgap of
quantum well structures.}} Phys. Rev. B Vol. 32, 1043.
\bibitem{Sun2} Wang A., Sun J., Zettl A.: {\sl\textit{The classification of self-adjoint boundary conditions: Separated, coupled , and mixed.}} Jour. Fun. Ana. (255) p 1554-1573, (2008).
\bibitem{Sun3} Wang A., Sun J., Zettl A.: {\sl\textit{Characterization of domains of self-adjoint ordinary differential operators}} Jou. Diff. Equ. (246) p 1600-1622, (2009).
\end{thebibliography}
\end{document}